\documentclass[aps,prb,amssymb,twocolumn,showpacs]{revtex4}

\usepackage{amsmath,amssymb}
\usepackage{amsfonts,amsthm}
\usepackage{epsfig}
\usepackage{graphicx}
\usepackage{dcolumn}
\usepackage{color}
\usepackage{bm}

\begin{document}
\title{Doped AB$_2$ Hubbard Chain: Spiral, Nagaoka and RVB States, Phase 
Separation and Luttinger Liquid Behavior}
\author{R.~R. Montenegro-Filho}
\email{rene@df.ufpe.br}
\author{M.~D. Coutinho-Filho}
\email{mdcf@ufpe.br}
\affiliation{Laborat\'orio de F\'{\i}sica Te\'orica e Computacional,
Departamento de F\'{\i}sica, Universidade Federal de Pernambuco, 50670-901, 
Recife-PE, Brazil}
\begin{abstract}
We present an extensive numerical study of the Hubbard model on the doped AB$_2$ 
chain, both in the weak coupling and the infinite-U limit. Due to the special 
unit cell topology, this system displays a rich variety of phases as function of 
hole doping ($\delta$) away from half-filling. Near half-filling, spiral states 
develop in the weak coupling regime, while Nagaoka itinerant ferromagnetism is 
observed in the infinite-U limit. For higher doping the system phase-separates 
before reaching a Mott insulating phase of short-range RVB states at 
$\delta=1/3$. Moreover, for $\delta>1/3$ we observe a crossover, which 
anticipates the Luttinger liquid behavior for $\delta > 2/3$.
\end{abstract}
\maketitle
\section{Introduction}
Low-dimensional strongly correlated electron systems have attracted great 
attention in the last two decades. The reason dates back to Anderson's proposal 
\cite{anderson_sce} that the $t-J$ version of the Hubbard model might carry the 
basic mechanisms underlying the high-Tc superconductivity observed in CuO$_2$ 
compounds. Despite that this remains an open issue, the above 
suggestion fertilized intensive investigations on many related fundamental 
topics, such us itinerant electron magnetism, Mott metal-insulator transitions 
and quantum critical phenomena. Amongst several features of interest, we mention 
the possibility of realization of spiral \cite{spiral}, Nagaoka \cite{nagaoka,tasaki,yso} 
and resonating-valence-bond (RVB) states \cite{rvb}, spatially separated phases 
\cite{emery,dagotto} and Luttinger liquid behavior \cite{luttinger}, which may present 
strong deviations from the Landau Fermi liquid theory. 

In this work, we report 
numerical results of the Hubbard model on the doped AB$_2$ chain away from half filling, which show that 
its special unit cell topology greatly enriches the phase diagram found in the 
doped standard linear chain. In fact, all features mentioned above are shown to 
be associated with well defined ground state (GS) phases of this doped chain. 
Doped AB$_2$-Hubbard chains were previously studied through Hartree-Fock, quantum monte carlo
 and exact diagonalization (ED) techniques both in the weak and strong coupling limits \cite{PRLMDCF}, including
also the $t-J$ model \cite{white} using density matrix renormalization group (DMRG) and
 recurrent variational Ans\"atzes, and the infinite-U limit \cite{JPSPW} using ED. In particular,
 these chains represent an alternative route to reaching two-dimensional quantum physics from
 one-dimensional systems \cite{Delgado,white}. At half filling the AB$_2$-Hubbard chain exhibits
 a quantum ferrimagnetic GS \cite{PRBTIAN,CELSO,PRLMDCF,PHYSARRMF}, whose magnetic excitations have been
 studied in detail both in the weak and strong coupling limits \cite{PHYSARRMF}, and 
 in the light of the quantum Heisenberg model \cite{PHYSARRMF,SW}. Further studies have considered the
 anisotropic \cite{CINV} and isotropic \cite{NLS} critical behavior of the AB$_2$-quantum-Heisenberg model, including
  its spherical version \cite{QSM}, and the statistical mechanics of the 
 AB$_2$-classical-Heisenberg model \cite{JPHYSA}.

\par On the experimental side, the AB$_2$ chain topology is of relevance to the understanding 
of the physics of some low-dimensional strongly correlated
electronic systems. One class is the line of trimer
clusters present in fosfates with formula A$_3$Cu$_3$(PO$_4$)$_4$, where A=Ca \cite{JBAnderson,Drillon,Belik,Matsuda},
 Sr \cite{Boukhari,Drillon,Belik,Matsuda} and Pb \cite{Eff,Belik,Matsuda}. The trimers
 have three Cu$^{2+}$ ($S=1/2$) paramagnetic ions antiferromagnetically coupled. Although
 the superexchange intertrimer interaction is much weaker than the intratrimer coupling, it proves
 sufficient to turn them bulk ferrimagnets. Another quasi-one-dimensional inorganic material closely associated
 with the ferrimagnetic phase of the AB$_2$ chain is the NiCu bimetallic chain \cite{NICU}. 
 These compounds display
 alternating Ni$^{2+}$ ($S=1$) and Cu$^{2+}$ ($S=1/2$) ions connected through suitable ligands in a
 line; and are modeled by the alternating spin-$\frac{1}{2}$/spin-1 antiferromagnetic Heisenberg chain \cite{spin1_2spin1}. We also would like to mention a more recently
 synthesized organic ferrimagnetic compound consisting of three $S=1/2$ paramagnetic radicals \cite{ORG}
 in its magnetic unit cell, as well as possible connections with the physics of the oxocuprates \cite{oxoc}.
  \par This paper is organized as follows: in Sec. II we introduce the model system and the numerical techniques used
  to calculate several quantities suitable to characterize the occurrence of distinct phases as function of doping and Coulomb
  coupling. In Sec. III, we discuss spiral and Nagaoka states at low hole doping, whose magnetic properties are 
  shown to exhibit very interesting features in the weak and infinite-U limit, respectively. In Sec. IV we show that
  for higher hole doping the system phase separates, before reaching a Mott insulating phase of short-range RVB states at 
  $\delta=1/3$. In Sec. V we discuss several features of the crossover region, which takes place before the Luttinger
  liquid behavior observed for $\delta>2/3$. Finally, in Sec. VI we present a summary and some conclusions concerning 
  the reported results.   
  
\section{Model description and Methods}
\begin{figure}[t]
\begin{center}
\epsfig{file=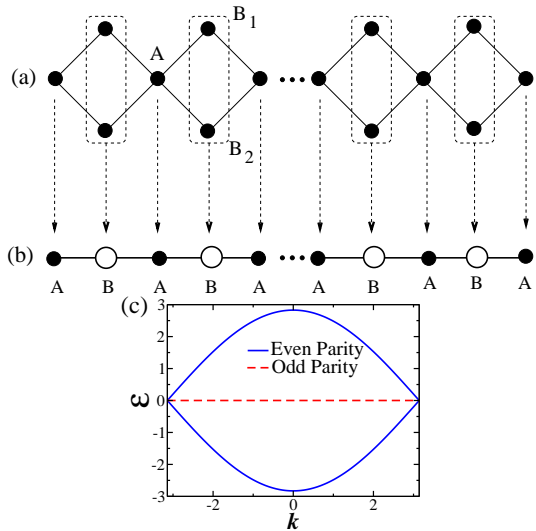,width=70mm,clip}\mbox{}
\epsfig{file=fig1c.eps,width=40mm,clip}
\caption{(Color online). (a) Illustration of the $AB_2$ chain showing $A$, $B_1$ and $B_2$ 
sites; (b) Illustration of the effective linear chain (ELC). (c) Electronic bands 
of the tight-binding model: two dispersive (continuous line) and one 
flat (dashed line).}
\label{Disp}
\end{center}
\end{figure}

The AB$_2$ chain
 is a bipartite lattice with three sites (named A, B$_1$ and B$_2$) per
unit cell, as illustrated in Fig. \ref{Disp}(a). The Hubbard Hamiltonian for a 
lattice with $N_c$ unit cells and $N$ sites reads:
\begin{equation}
H=-t\sqrt{2}\sum_{l=1,\sigma}^{N_c}[b_{l\sigma}^{\dagger}
(A_{l\sigma}+A_{l+1,\sigma})+\mbox{H.c.}]+U\sum_{i=1}^Nn_{i\uparrow}n_{i\downarrow},\end{equation}
where $A_{l\sigma}^{\dagger}$ and 
$b_{l\sigma}^{\dagger}=\frac{1}{\sqrt{2}}(B_{1,l\sigma}^\dagger+B_{2,l\sigma}^\dagger)$ are the creation operators of an electron with spin $\sigma$ at site A 
and in a bonding state between sites $B_1$ and $B_2$ of the cell $l$, 
respectively, $t(\equiv 1)$ is the hopping amplitude and $U$ is the Coulomb 
coupling. For $U=\infty$, double occupancy is completely excluded and the 
Hamiltonian takes the 
form:
\begin{equation}
H=-t\sqrt{2}\sum_{l=1,\sigma}^{N_c}P_G[b_{l\sigma}^{\dagger
}(A_{l\sigma}+A_{l+1,\sigma})+\mbox{H.c.}]P_G,
\end{equation}
where 
$P_{G}=\prod_{i}(1-n_{i\uparrow}n_{i\downarrow})$ is the Gutzwiller projector 
operator. The model is invariant under the interchange of the $B$ sites of the 
same cell, a symmetry that implies in a well defined local parity ($p_l=\pm1$) for 
the GS wave function. As a result, in computing some quantities we find it convenient
 to use the {\it effective linear 
chain} (ELC) generated by the map illustrated in Figs. \ref{Disp}(a) and 
\ref{Disp}(b), i. e., any quantity $X_{B,l}$ associated with a $B$ site at cell $l$ of 
the ELC is given by $X_{B_1,l}+X_{B_2,l}$. This mapping does not change the physical content
of the GS and excited states, being used only to expose in a more
clear fashion some properties of these states.
\par In 
the tight-binding description ($U=0$) this model presents three bands \cite{PRLMDCF}: one 
flat with $N_c$ odd parity states [antibonding orbitals, 
$a^{\dagger}_{l\sigma}=\frac{1}{\sqrt{2}}(B^{\dagger}_{1,l\sigma}-B^{\dagger}_{2
,l\sigma}$)] and energy $\epsilon=0$; and two dispersive 
branches,
\begin{equation}
\epsilon_{\pm}(k)=\pm 2\sqrt{2}\cos(k/2), 
\label{disp}
\end{equation}
with $k=2\pi l/N_c$, $l=0,1,2,...,N_c-1$,
 built from A
sites and bonding (even parity) orbitals, as shown in Fig. \ref{Disp}(c). At half filling 
($N_e=N$, where $N_e$ is the number of electrons) the GS total spin $S_g$ is 
degenerate, with $S_g$ ranging from the minimum value ($0$ or $1/2$) to 
$S_g=|N_B-N_A|/2$, where $N_A(N_B)$ is the number of sites in the A (B) 
sublattice. As proved by Lieb \cite{PRLLIEB} the Coulomb repulsion lifts 
this huge degeneracy and selects the 
\begin{equation}
S_g=|N_B-N_A|/2\equiv S_{Lieb} 
\end{equation}
ground state for any
finite $U$, giving rise to a ferrimagnetic GS \cite{PRBTIAN,PRLMDCF,PHYSARRMF}.
 
 On the other hand, for $U=\infty$, one hole ($N_e=N-1$) and periodic
boundary conditions (BC's), the system satisfies the requirements of Nagaoka's 
theorem for saturated ferromagnetism \cite{PRLMDCF,JPSPW}. For Nagaoka 
ferromagnetism and Lieb ferrimagnetism the GS is homogeneous in parity with 
$p_l=-1$ for any cell $l$. Due to this symmetry, the spectrum of the AB$_2$ chain in the Heisenberg limit ($U>>t$, $N_e=N$)
at the sector $p=-1$  is identical to that
of the alternating Heisenberg spin-$\frac{1}{2}$/spin-1 chain \cite{spin1_2spin1}.

\begin{figure}[b]
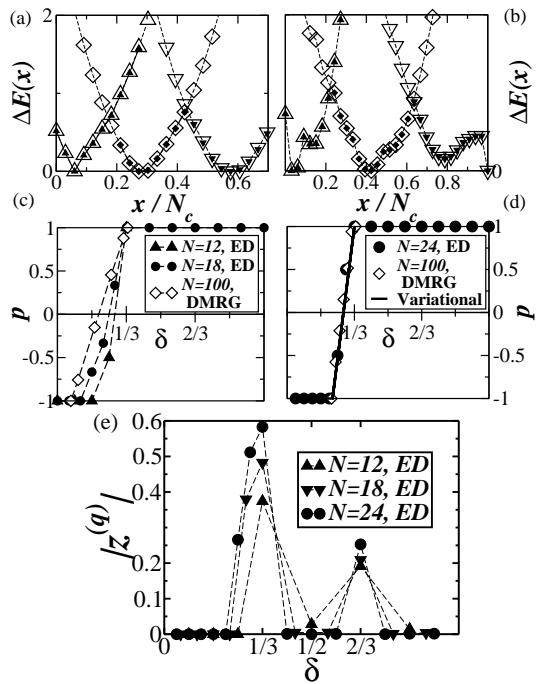

\begin{center}
\epsfig{file=fig2abcd.eps,width=70mm,clip}
\epsfig{file=fig2e.eps,width=50mm,clip}
\caption{DMRG results for the energy difference between 
the lowest energy at the symmetry sector $(-)^x(+)^{N_c-x}$ and the GS energy for 
$N=100$ ($N_c=33$) for (a) $U=2$ at $\delta=0.32$ (triangle up), 
$\delta=0.26$ (diamond) and  $\delta=0.18$ (triangle down); (b) $U=\infty$ 
at $\delta=0.32$ (triangle up), $\delta=0.28$ (triangle up) and 
 $\delta=0.24$ (triangle down), taking 108 states per block (open symbols) and 
216 states per block (filled symbols). Average local parity $p$ as function of $\delta$
for (c) $U=2$ and (d) $U=\infty$. (e) ED results for $|z^{(q)}|$. Dashed 
lines are guides to the eye.} \label{pdEz}
\end{center}
\end{figure}

\begin{figure}[t]
\begin{center}\epsfig{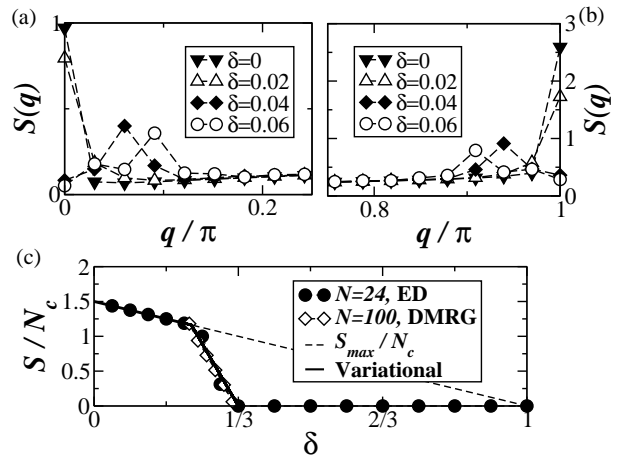}
\caption{ (a) and (b): Magnetic structure factor for $U=2$ and $N=100$ in the underdoped 
region. (c) Total spin per cell $S/N_c$ as function of $\delta$ for $U=\infty$.}
\label{sqsefl}
\end{center}
\end{figure}

Here we focus on the effect of hole doping, $\delta=1-(N_e/N)$, 
both in the weak coupling and the infinite-U limit, using exact diagonalization 
(ED) through the Lanczos algorithm for closed BC's and DMRG for open BC's \cite{dmrg1}.
In the ED procedure, the BC's are such to minimize the energy,
except for $U=2$ and $\delta\leq 1/3$ [Fig. \ref{pdEz}(c)] in which the BC's (periodic or antiperiodic) are such
that the Fermi wave vector $k_F$ in the thermodynamic limit is included in the set of wave vectors for the
finite system \cite{julien}.
We used finite size DMRG for open chains with
A sites in its extrema, keeping 364 to 546
states per block in the last sweep. The maximum discarded weight in the last 
sweep was typically $\thicksim 10^{-7}$, except for odd phases and $U=2$, where the discarded weight was $\thicksim 10^{-5}$.
In the DMRG calculations we treated $B_1$ and $B_2$ as a composite site
with 9 states for $U=\infty$ and 16 states for $U=2$.
However, by considering the parity symmetry, we can decompose
this supersite into the two possible symmetry sectors $+1$ and $-1$
. Within this scheme, we have considered all parity symmetry
 sectors of the form 
$(-)^x(+)^{N_c-x}$, with $x$ contiguous cells of odd parity 
in one side of the open chain and $N_c-x$ contiguous cells of even parity 
in the other. In addition, we have verified the stability of this phase separation against the formation of a mixed phase 
composed of smaller domains. The energy is studied as function of $x$ for
increasing number of states kept per block in order to localize the value of
$x$ for which the energy is minimum, as shown in Figs. \ref{pdEz}(a) and \ref{pdEz}(b). 
The phase-separated boundaries are thus determined by the limiting dopings
for which an inhomogeneous phase (non-uniform parities) is observed.
We have also developed a simple variational approach for $U=\infty$ and $\delta\leq 1/3$, 
which is explained in detail in Appendix A. The results calculated using this approach are shown in Figs. 
\ref{pdEz}(d) and \ref{sqsefl}(c).  

In Fig. \ref{pdEz}(c) ($U=2$) and Fig. \ref{pdEz}(d) ($U=\infty$) 
we present the average parity, 
 \begin{equation}
 p\equiv\frac{1}{N_{c}}\sum_{l=1}^{N_{c}} p_{l},
 \end{equation} 
 as function of doping, computed using the above-mentioned methods.
 In both regimes, we observe the occurrence of an homogeneous phase near half 
filling with $p=-1$. For higher doping, i. e., 
$\delta_{\mbox{PS}}(U)<\delta<(1/3)$ [$\delta_{\mbox{PS}}(2)\approx 0.07$  and 
$\delta_{\mbox{PS}}(\infty)\approx 0.22]$ the system phase separates in one 
region  with odd parity cells and the other with even ones. For $\delta \geq 
1/3$ the GS is homogeneous with $p=1$. 

In order to present an overview of 
the conducting properties of the $AB_2$ chain phases in the infinite-U limit, 
we display in Fig. \ref{pdEz}(e) the quantity \cite{zq}
\begin{equation}
 |z^{(q)}|=|\langle \exp{(\frac{2\pi q i}{L}\sum_j{x_j})} \rangle|,
 \end{equation}
 calculated in the ELC using ED,
 where $L=2N_c$, $x_j=jn_j$, 
 $n_j$ is the electron density at site $j$ and $q$ is such that 
$\frac{N_e}{L}=\frac{p}{q}$, with $p$ and $q$ co-primes. The phase of $z^{(q)}$ 
corresponds to the GS expectation value of the position operator, while its modulus 
defines the localization length; in an insulator, 
$|z^{(q)}|\rightarrow 1$, as $L\rightarrow \infty$, while in a conductor, 
$|z^{(q)}|\rightarrow 0$, for closed boundary 
conditions \cite{zq}. The increase of $|z^{(q)}|$ with system size for $\delta=2/3\mbox{ and }1/3$, as well as in the phase 
separated region, are evidences of insulating phases at these dopings. 
These conclusions will be better fundamented by studying the Drude weight using ED and the charge gap
 for larger systems with DMRG. 

\section{Spiral states and Saturated ferromagnetism}

In Figs. \ref{sqsefl}(a) and \ref{sqsefl}(b) we display
the magnetic structure factor
\begin{equation}
S(q)=\frac{1}{S_{Lieb}(S_{Lieb}+1)}\sum_{l,m}^{2Nc+1}e^{iq(l-m)}\langle\mathbf{S
}_l\cdot\mathbf{S}_m\rangle,
\end{equation}
calculated at $S^z=0$ and $U=2$ using DMRG  
for the ELC. First, notice the presence of peaks at $q=0$
 and $q=\pi$ revealing the ferrimagnetic order at half filling. These peaks 
sustain up to two holes ($\delta=0.02$); however, it is not clear whether the 
ferrimagnetic phase is robust against doping in the thermodynamic 
limit. Indeed, by increasing the hole doping, spiral peaks at $\delta$-dependent 
positions appear near $q=0$ and $q=\pi$. The analysis of the charge gap, 
\begin{equation}
\Delta_c=E(N_e+1)+E(N_e-1)-2E(N_e),
\end{equation}
suggests that
these states are metallic, in opposition to the Mott insulating ferrimagnetic 
state at $\delta=0$. It is worth mentioning that the occurrence of spiral phases in oxocuprates 
has been a challenging and topical subject \cite{oxoc}.
 
 In Fig. \ref{sqsefl}(c) we present  the GS total spin as function of doping for 
$U=\infty$. For $\delta<\delta_{\mbox{PS}}(\infty)$ itinerant saturated 
ferromagnetism due to hole kinematics (Nagaoka mechanism) is observed. It is interesting to notice
that our estimate for the upper hole density ($\gtrsim 0.2$) beyond which Nagaoka ferromagnetism is 
unstable is in very good agreement with similar predictions for ladders \cite{troyer,1D_F} and
the square lattice \cite{2D_F}.

We have 
also considered the presence of an Aharonov-Bohn flux $\Phi$ for a closed 
chain through the gauge transformation:
\begin{equation}
 \left\{\begin{array}{ccc}
        b_{l\sigma} & \rightarrow & b_{l\sigma}\mbox{e}^{2\pi \Phi l/N_c};\\
        A_{l\sigma} & \rightarrow & A_{l\sigma}\mbox{e}^{2\pi \Phi l/N_c},
        \end{array}
\right .
\end{equation}
\begin{figure}[t]
\begin{center}
\epsfig{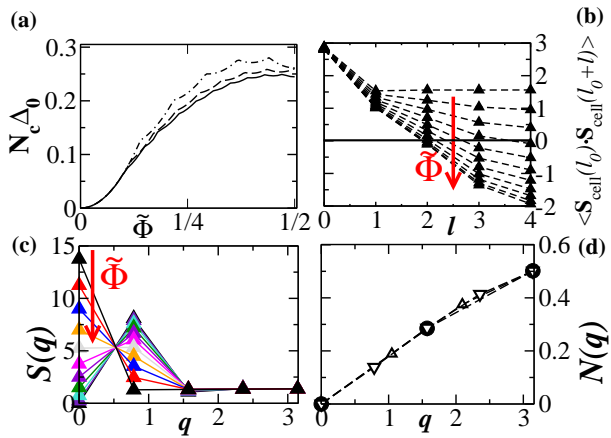}
\caption{(Color online). ED results of (a) $N_{c}$ times the energy gap $\Delta_0$ between 
the saturated ferromagnetism energy 
($\Phi=\Phi_F$) and the lowest energy state for an Aharanov Bohn flux $\Phi$
as function of $\widetilde{\Phi}=|\Phi-\Phi_{F}|$
at $\delta=1/6$ and $N_c=$ 4 (dashed-dot line), 6 (dashed line) and 8 (solid 
line). ED calculation for the $\widetilde{\Phi}$ dependent behavior of the (b) 
Spin correlation function $\langle 
\mathbf{S}_{\mbox{cell}}(l_0)\cdot\mathbf{S}_{\mbox{cell}}(l_0+l)\rangle$ between 
the cell spins as function of $l$ and (c) magnetic structure factor as function 
of lattice wave vector $q$ at $\delta=1/6$. (d) Charge structure factor calculated at the lowest energy state
for any $\widetilde{\Phi}$ at $\delta=1/6$ for $N_c=$ 4 
($\bullet$), 6 ($\blacktriangle$) and 8 ($\triangledown$).}\label{d0scn}
\end{center}
\end{figure}
with $\Phi_0=hc/e \equiv 1$. The flux variation is equivalent to a change 
in the boundary condition: $\Phi=0$ represents periodic and $\Phi=1/2$ 
antiperiodic boundary conditions. In Fig. \ref{d0scn}(a) we present the 
dependence of the energy gap $\Delta_0$ between the lowest energy state for a flux 
$\Phi$ and that for saturated ferromagnetism ($\Phi\equiv \Phi_F$) as function 
of $\widetilde{\Phi}=|\Phi-\Phi_{F}|$ at $\delta=1/6$. We have identified many 
level crossings in this curve. In fact, as the flux increases from $\Phi_F$, the 
total spin decreases from the maximum value, $S=N_e/2$, to the minimum value $S=0$ 
($S=1/2$) for $N_e$ even (odd), a behavior also observed in the square lattice 
\cite{kusakabe}. Notice that $N_c\Delta_0$ tends to saturation with system size,
 indicating that the level spacings decrease with $1/N_c$. 
 These results suggest that the thermodynamic GS displays spontaneously SU(2) 
  symmetry breaking as a result of
 an ergodic combination of infinitely many states ($N_c\rightarrow \infty$),
  including the singlet spiral state   
 \cite{koma_e_tasaki}. In Figs. \ref{d0scn}(b) and \ref{d0scn}(c) we present the 
spin correlation function between cell spins 
$\mathbf{S}_{cell}(l)=\mathbf{S}_A(l)+\mathbf{S}_{B_1}(l)+\mathbf{S}_{B_2}(l)$ 
and the magnetic structure factor 
\begin{equation}
S(q)=\frac{1}{N_c}\sum_{\langle l,m \rangle} 
e^{iq(l-m)}\langle\mathbf{S}_{cell}(l)\cdot \mathbf{S}_{cell}(m) 
\rangle
\end{equation}
as function of distance 
$l$ and wave vector $q=2\pi l/N_c$, $l=0,...,N_c$, respectively.
As we can observe, the saturated ferromagnetic and the spiral singlet states are 
adiabatically connected, such that all states contributing to the thermodynamic 
GS exhibit long-range ordering. 
In particular, as the flux increases from $\Phi_F$ the peak of 
$S(q)$ at $q=0$ (saturated ferromagnetism) steadily decreases,
while the spiral state peak at $q=2\pi/N_c$ increases.  
We noted also that the charge structure factor
\begin{equation}
N(q)=\frac{1}{N_c}\sum_{\langle l,m \rangle} e^{iq(l-m)}\langle \Delta n_l 
\Delta n_m \rangle,
\end{equation}
where $\Delta n_l=n_l-\langle n_l \rangle$ and $n_l$ is the electron occupation 
number at cell $l$, is not affected by the flux variation and displays a peak at 
$2k_F=\pi$ [Fig. \ref{d0scn}(d)], where $k_F$ is the tight-binding {\it 
spinless} Fermi wave vector \cite{PRLMDCF},  with $k_F=3\pi \delta$, $\delta\leq 
1/3$.
\section{Phase separation and RVB states}

\begin{figure}[b]
\begin{center}
\epsfig{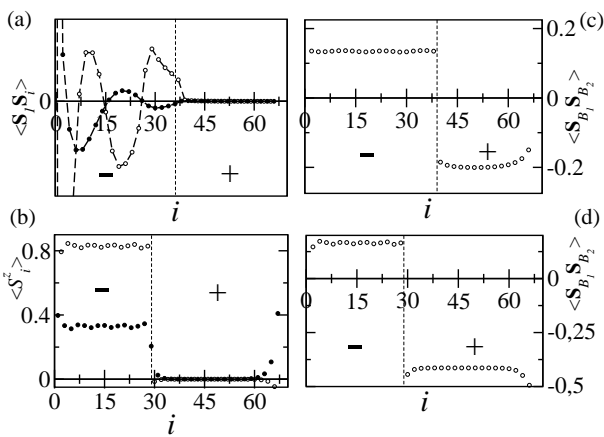}
\caption{GS properties at $\delta=0.18$ ($U=2$) and $\delta=0.28$ ($U=\infty$) 
for $N=100$ using DMRG. (a) Spin correlation function $\langle 
\mathbf{S}_{1}\cdot\mathbf{S}_{i}\rangle$ for $U=2$. (b) Expectation value of 
$S^z_i$ for $U=\infty$ in the sector $S^z=S_g$. Spin correlation function 
$\langle \mathbf{S}_{B_1}\cdot\mathbf{S}_{B_2}\rangle_i$ for (c) $U=2$ and (d) 
$U=\infty$. $-$($+$) indicates odd (even) local parity. Dashed lines are guides 
to the eye.}
\label{cszb1b2}
\end{center}
\end{figure}

\begin{figure}[t]
\begin{center}
\epsfig{file=fig6ab.eps,width=80mm,clip}
\epsfig{file=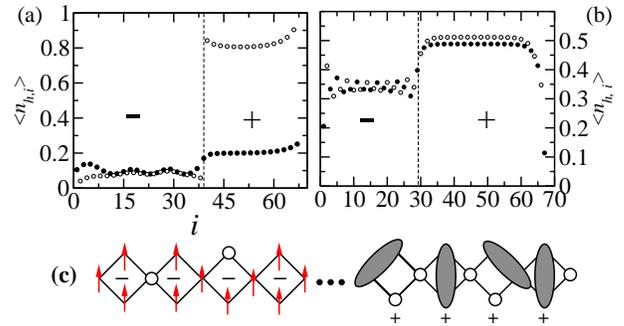,width=70mm,clip}
\caption{(Color online). GS properties at $\delta=0.18$ 
($U=2$) and $\delta=0.28$ ($U=\infty$) for $N=100$ using DMRG. Expectation value 
of $n_{h,i}$ for (a) $U=2$ and (b) $U=\infty$. Effective linear chain 
notation:($\bullet$) identifies $A$ sites and ($\circ$) $B_1+B_2$ at the same 
cell. (c) Illustration of the GS for $U=\infty$ in the phase-separated regime: 
singlet bonds are represented by ellipses and holes by circles. $-$($+$) 
indicates odd (even) local parity. Dashed lines are guides to the 
eye.}
\label{nhs}
\end{center}
\end{figure}

In the phase-separated regime the charge compressibility
diverges following the linear dependence of the energy with doping. In Figs. 
\ref{cszb1b2} and \ref{nhs} we present some properties of the GS
in this regime calculated through DMRG for the ELC. First we 
notice that all these properties clearly exhibit some modulation on the same 
sublattice in the {\it metallic} odd parity region due to charge itinerancy. In 
particular, this modulation is stronger in the $U=2$ spiral phase as evidenced by 
the correlation function $\langle\mathbf{S}_1\cdot\mathbf{S}_i \rangle$ shown in 
Fig. \ref{cszb1b2}(a), but also noticed in the itinerant Nagaoka phase 
($U=\infty$) as manifested by the site magnetization $\langle S^z_i\rangle$ shown 
in Fig. \ref{cszb1b2}(c). On the other hand, in the {\it insulating} even parity 
phase a flat behavior is observed, except for boundary and interface effects. 
These paramagnetic phases [see Figs. \ref{cszb1b2}(b) and \ref{cszb1b2}(d)] 
are characterized by strong singlet correlations between spins at sites $B_1$ and 
$B_2$ at the same cell, i. e., $\langle\mathbf{S}_{B_1}\cdot\mathbf{S}_{B_2} 
\rangle\approx -0.20$ $(\approx -0.41)$ for $U=2$ $(=\infty)$, as shown in Figs. 
\ref{cszb1b2}(b) and \ref{cszb1b2}(d). In contrast, in the metallic phase this 
correlation varies very little with $U$ and indicates robust triplet 
correlations, i. e., $\langle\mathbf{S}_{B_1}\cdot\mathbf{S}_{B_2} 
\rangle\approx 0.13$ $(\approx 0.16)$ for $U=2$ $(=\infty)$. Notice that in the 
absence of hole hopping, even when restricted to a cell as in the insulating 
phase, the value of $\langle\mathbf{S}_{B_1}\cdot\mathbf{S}_{B_2}\rangle$ in a 
singlet (triplet) state should be $-0.75$ ($0.25$). The hole density $\langle 
n_{h,i}\rangle$ is shown in Figs. \ref{nhs}(a) and \ref{nhs}(b). In the odd 
parity metallic phase, holes do not occupy antibonding orbitals, whereas in the 
even parity insulating phase these orbitals are accessible for them. Therefore, 
in the first case the hole densities at sites $A$ and $B_1+B_2$ are very similar. 
This may also occur in the second case if double occupancy is excluded 
($U=\infty$). An illustration of the phase-separated regime for $U=\infty$ is 
shown in Fig. \ref{nhs}(c). In this coupling limit, unsaturated ferromagnetism was suggested 
to occur in ladders \cite{1D_F} and the square lattice \cite{2D_F} as an 
intermediate phase between saturated ferromagnetism and paramagnetism as 
function of doping. However, in the context of the $t-J$ model the situation is more complex
and predictions of phase separation, both for ladders \cite{troyer} and the square lattice \cite{emery,WhiteScalapino,Altshuler}, and stripe formation for the square lattice \cite{WhiteScalapino} have been reported.

\begin{figure}[t]
\begin{center}
\epsfig{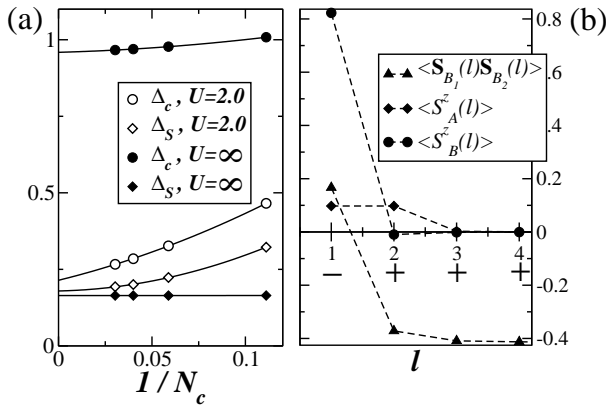}
\caption{DMRG results for the (a) size dependence of the charge ($\Delta_c$)
 and spin ($\Delta_S$) gaps as function of $1/N_c$ at $\delta=1/3$:
 solid lines are polynomial fittings. ED results for the (b) expectation values 
of  $S^z_A(l)$, $S^z_B(l)$ and the correlation function $\langle 
\mathbf{S}_{B_1}(l)\cdot\mathbf{S}_{B_2}(l)\rangle$  at spin sector $S^z=1$ as 
function of cell number $l$. The $\pm$ signs below the horizontal axis in (b) 
indicate the cell parity. }
\label{dsdcsz13}
\end{center}
\end{figure}

At $\delta=1/3$, i. e., one hole per $A$ 
site for open BC's using DMRG \cite{white}, the GS has even parity and is fully 
dominated by the Mott insulating phase (even parity) illustrated in Fig. 
\ref{nhs}(c) for $U=\infty$. The charge gap $\Delta_c=\mu_+-\mu_-$, where 
$\mu_{+}=[E(N_e+\Delta N_e)-E(N_e)]/\Delta N_e$, $\Delta N_e>0$ ($\Delta 
N_e/N\rightarrow 0$), and $\mu_{-}=E(N_e)-E(N_e-1)$, must be calculated with 
care. First, notice that adding electrons to $\delta=1/3$ places the system in 
the phase-separated (inhomogeneous) region where the chemical potential $\mu$ is 
flat. Indeed, by comparing results using DMRG and ED calculations for 
$U=\infty$, for which $\Delta_c$ presents little finite size corrections [Fig. 
\ref{dsdcsz13}(a)], we concluded that boundary effects are minimized by taking $\Delta N_e=2$ 
and placing the symmetry inverted cells at the chain center. We thus find [Fig. 
\ref{dsdcsz13}(a)] $\Delta_c\approx 0.21$ ($\approx 0.96$) for $U=2$ 
($=\infty$). This problem is absent in the case of hole doping since the phase 
is homogeneous. The extrapolated spin gap, 
\begin{equation}
\Delta_S=E(S=1)-E(S=0),
\end{equation} 
characterized by symmetry 
inversion of a cell at the chain center, is also shown in Fig. \ref{dsdcsz13}(a) 
for $U=2$ ($\Delta_S\approx 0.18$) and $U=\infty$ ($\Delta_S\approx 0.16$), with 
the spin gap at $U=\infty$ presenting little finite size dependence. It is a quite massive 
excitation with the magnon localized at the odd symmetry cell, mostly at the B 
sites, as shown in Fig. \ref{dsdcsz13}(b). In this context, Sierra {\it et al.} \cite{white} found 
$\Delta_S\approx 0.27$ using the $t-J$ model ($J=4t^2/U$) for $J=0.35t$, i. e., 
$U\approx 11.43$. We have confirmed this result by studying the $U$ dependence 
of $\Delta_S$ using ED.
\begin{figure}[t]
\begin{center}
\epsfig{file=fig8abc.eps,width=65mm,clip}
\epsfig{file=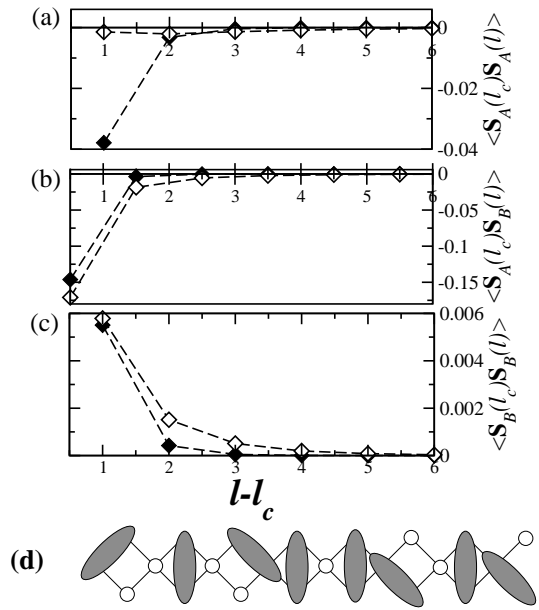,width=70mm,clip}
\caption{Spin correlation functions (a) $\langle \mathbf{S}_A(l_c)\cdot\mathbf{S}_A(l)\rangle$,
 (b) $\langle \mathbf{S}_A(l_c)\cdot\mathbf{S}_B(l)\rangle$ and (c) 
 $\langle \mathbf{S}_B(l_c)\cdot\mathbf{S}_B(l)\rangle$ as function of $l-l_c$ in the
 ELC at $\delta=1/3$ for $N_c=33$ using DMRG; in the above expressions $l_c$ denotes the central cell;
 ($\lozenge$) refers 
to $U=2$ and ($\blacklozenge$) to $U=\infty$. Dashed lines are guides to the eye. 
(d) Illustration of the GS at $\delta=1/3$, singlet bonds are represented by 
ellipses and holes by circles.}
\label{cor13}
\end{center}
\end{figure}
In Fig. \ref{cor13} we show that the spin correlation functions at $\delta=1/3$,
calculated using DMRG, present a fast decay and can be fitted with the exponential form 
$\exp{[-(l-l_c)/\xi)]}$, where $\xi$ is the correlation length, $l$ is the cell index in the
ELC and $l_c$ denotes the central cell of the system. This behavior is expected from 
the presence of a finite spin gap. The values of $\xi$ for the correlations 
$\langle \mathbf{S}_A(l_c)\cdot\mathbf{S}_A(l)\rangle$,
$\langle \mathbf{S}_A(l_c)\cdot\mathbf{S}_B(l)\rangle$ and $\langle \mathbf{S}_B(l_c)\cdot\mathbf{S}_B(l)\rangle$
are $\approx$ 0.4 (2.2), 0.25 (0.45) and 0.39 (0.75), respectively, for $U=\infty$ ($U=2$), with $l_c$ denoting
 the central cell. Thus, 
except for the correlation
$\langle \mathbf{S}_A(l_c)\cdot\mathbf{S}_A(l)\rangle$ at $U=2$, the correlation length is extremely short 
with spins correlated only within a cell.
Further, the calculated bulk values of $\langle \mathbf{S}_{B_1}\cdot\mathbf{S}_{B_2} \rangle$ at $\delta=1/3$
are in very good agreement with those in the even phase of the separated region shown in 
Figs. \ref{cszb1b2}(b) and \ref{cszb1b2}(d). The above results support a 
short-range-RVB (SR-RVB) \cite{sr_rvb} state for the GS at $\delta=1/3$, as illustrated in Fig. \ref{cor13}(d).  
 In this context, Sierra {\it et al.} \cite{white} reached similar conclusions using the $t-J$ model on the AB$_2$ chain,
 while Giesekus has proved \cite{giesekus} that a SR-RVB state is the GS of a 
non-bipartite lattice with the same local symmetry but a different hopping 
pattern.
\section{Luttinger Liquid Behavior}
\begin{figure}[t]
\begin{center}
\epsfig{file=fig9a.eps,width=60mm,clip}
\epsfig{file=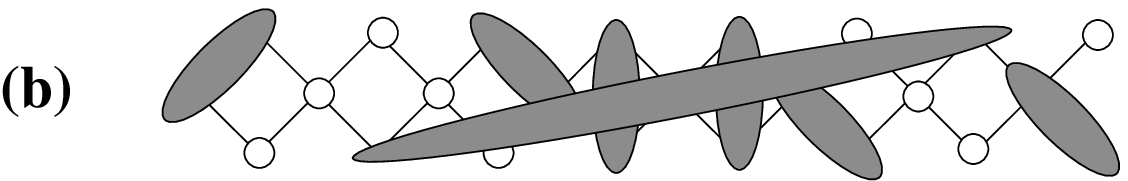,width=70mm,clip}
\epsfig{file=fig9cd.eps,width=80mm,clip}
\caption{(a) ED results for the indicated spin correlation functions as function
of cell index $l$, with $\mathbf{S}_B\equiv \mathbf{S}_{B_1}+\mathbf{S}_{B_2}$.
(b) Illustration of the GS at $\delta=1/3$ doped
with two holes: singlet bonds are represented by ellipses and holes by circles. 
(c) ED results for the
spin correlation functions between $B$ sites at the same cell $\langle 
\mathbf{S}_{B_1}(l_0)\cdot \mathbf{S}_{B_2}(l_0)\rangle$, electron densities at 
$A$ sites $\langle n_A(l_0)\rangle $ and at $B\equiv B_1+B_2$ sites 
$\langle n_B(l_0)\rangle\equiv \langle n_{B_1}(l_0)+n_{B_2}(l_0)\rangle$. (d) ED results for the 
indicated nearest-neighbor spin correlation functions as function of $\delta$,
with $\mathbf{S}_B\equiv \mathbf{S}_{B_1}+\mathbf{S}_{B_2}$.
In (a), (c) and (d) $l_0$ denotes an arbitrary cell.}
\label{cor13dopcross}
\end{center}
\end{figure}
\begin{figure}
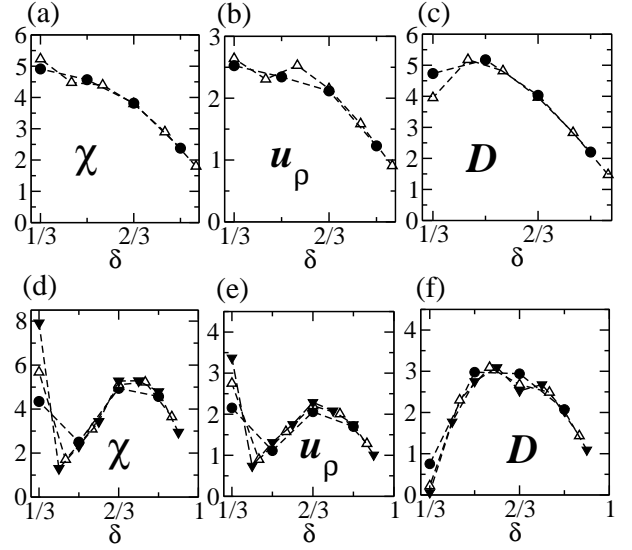

\begin{center}
\epsfig{file=fig10abc.eps,width=80mm,clip}
\epsfig{file=fig10def.eps,width=80mm,clip}
\caption{ED results for the charge susceptibility $\chi$, the charge excitation velocity $u_\rho$ and 
the Drude weight $D$, for $U=2$ [(a),(b) and (c)] and $U=\infty$ [(d), (e) and (f)],
and $N_c=$ 4 ($\bullet$), 6 ($\triangle$) and 8 ($\blacktriangledown$).
}
\label{csiud}
\end{center}
\end{figure}
\begin{figure}[t]
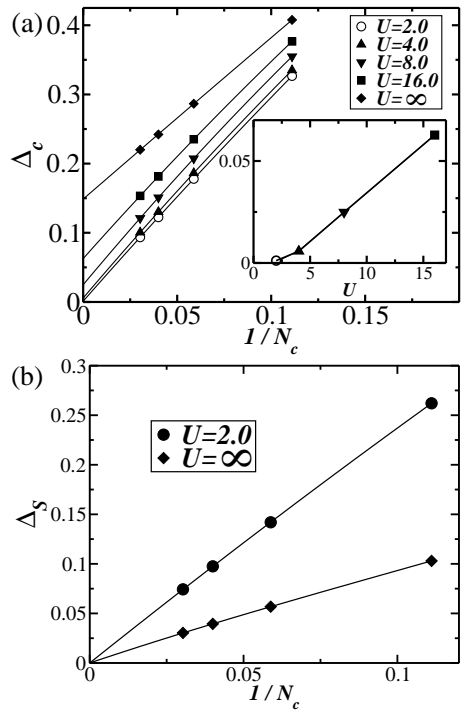

\begin{center}
\epsfig{file=fig11a.eps,width=60mm,clip}
\epsfig{file=fig11b.eps,width=60mm,clip}
\caption{DMRG results for the (a) Charge gap $\Delta_c$ as function of $1/N_c$ 
at $\delta=2/3$ using DMRG; the inset 
presents extrapolated values of the charge gap as function of $U$
. DMRG calculation of the (b) Spin gap
$\Delta_S$ as function of $1/N_c$ for $U=2$ ($\bullet$)
and $U=\infty$ ($\blacklozenge$). Solid lines are polynomial fittings, except
in the inset of (a), where we have used an essential singularity form as explained in the text.
 }
\label{dsdc23}
\end{center}
\end{figure}
We now focus on the behavior of the system for $1/3<\delta<1$ by
considering a chain with closed boundary conditions and $N_c=8$ for $U=\infty$ using
ED.
The first noticeable feature is the behavior of the spin 
correlation functions after doping the $\delta=1/3$ GS with
two holes. The value of $\langle \mathbf{S}_{B_1}(l_0)\cdot 
\mathbf{S}_{B_2}(l_0)\rangle$ (where $l_0$ denotes an arbitrary cell)
changes from -0.41 to -0.28. This variation can be understood
by considering that the two holes added to the system break two singlet bonds 
and reside predominately at B sites. In this picture the correlation function 
would amounts to $\frac{N_c-2}{N_c}(-0.41)\approx -0.31$, which is close to -0.28. 
Furthermore, the spin correlation functions shown in Fig. \ref{cor13dopcross}(a) 
evidence the formation of long ranged bonds between electrons on $B$ sites, while 
the other correlations remain short ranged, as in the $\delta=1/3$ ground 
state. This fact indicates that the electrons picked from the SR-RVB by hole 
doping are antiferromagnetically coupled and delocalized through the system, 
as illustrated in Fig. \ref{cor13dopcross}(b). 
In order to describe the system behavior for finite dopings, we display in Fig. 
\ref{cor13dopcross}(c) the correlation function $\langle \mathbf{S}_{B_1}(l_0)\cdot \mathbf{S}_{B_2}(l_0)\rangle$ 
 and electronic densities as function of $\delta$. 
Notice that for $1/3<\delta<2/3$ the electronic 
density at $A$ sites is almost fixed, while that at $B$ sites are monotonically depopulated. 
As a 
consequence, $\langle \mathbf{S}_{B_1}(l_0)\cdot \mathbf{S}_{B_2}(l_0)\rangle$ 
continuously vanishes as the doping increases. Moreover, in Fig. 
\ref{cor13dopcross}(d) we show the relevant nearest-neighbor spin correlation functions. These 
correlations display quite different magnitudes at $\delta=1/3$, but their values approach each other
for $\delta>2/3$. We thus consider the doping interval $1/3<\delta<2/3$ as a {\it crossover
region}, where doping starts to build the Luttinger liquid which is fully
established for $\delta>2/3$. 

We have also calculated the charge compressibility $\kappa$ through
\begin{equation}
\chi=\frac{1}{n_0^2\kappa}=\frac{V}{4}[E(N_e+2)+E(N_e-2)-2E(N_e)],\label{ll2}
\end{equation}
where $V$ is the volume and $n_0=\frac{N_e}{V}$ is the
electronic density; the charge excitation velocity
\begin{equation}
u_\rho=\frac{E(\Delta k,S=0)-E_{GS}}{\Delta k},
\label{ll3}
\end{equation}
with $\Delta k=2\pi/L$ and $L$ the system length; and the Drude weight 
\begin{equation}
D=\frac{L}{4\pi}\left[\frac{\partial^2 E(\Phi)}{\partial 
\Phi^2}\right]_{\Phi_{min}},\label{ll4}
\end{equation}
where $\Phi_{min}$ is the flux value that minimizes the energy \cite{fye}.
In an insulating phase these quantities satisfy the limits below
\begin{equation}
\lim_{N_c\rightarrow \infty}\left \{
\begin{array}{ccc}
\chi & = & \infty;\\
u_\rho & = & \infty;\\
D & = & 0,
\end{array}
\right .    
\end{equation}
while for a metal $\chi$, $u_\rho$ and $D$ are finite. As shown in Fig. \ref{csiud},
at $\delta=1/3$, $\chi$ and $u_\rho$ increases, while $D$ 
decreases with system size
for both $U=2$ and $U=\infty$, although the insulating character is 
better evidenced for $U=\infty$ due to its sizable charge gap, 
as shown in Fig. \ref{dsdcsz13}(a). At the other commensurate density, $\delta=2/3$, we 
can see the signals of an insulating phase for $U=\infty$, while for $U=2$ we does not 
observe any especial behavior. In order to clarify this point, we have used DMRG to 
study the size dependence of the charge gap for larger systems at this doping. 
For a finite open chain, the occupation of two holes per cell tends to 
$\delta=2/3$ in the thermodynamic limit. In Fig. 
\ref{dsdc23}(a), we can clearly observe that for $U=\infty$ the system is in a Mott insulating phase with 
$\Delta_c\approx 0.15$; however, the gap for $U=2$ is extremely 
small. In order to better understand the U-dependence of this gap, we have also calculated $\Delta_c$ for 
intermediate values of $U$, as also shown in Fig. \ref{dsdc23}(a). 
In the inset of Fig. \ref{dsdc23}(a) we have fit $\Delta_c(U)$ using an expression similar
 to the limiting behavior of the charge
gap as $U\rightarrow 0$ of the Lieb-Wu solution for a linear chain at half filling \cite{liebwu}: 
$U^{a}\exp{(-b/x)}$, in which $a\approx 0.61$ and $b\approx 7.95$ are fitting parameters. 
Notice, however, that contrary to the Lieb-Wu solution \cite{liebwu}, $\Delta_c$ saturates to a finite
value ($\approx 0.15$) for $U=\infty$. On the other hand, similarly to the linear chain at half filling \cite{liebwu},
 the data shown in Fig. \ref{dsdc23}(b) indicates the absence of spin gap at $\delta=2/3$ in the thermodynamic limit
 for both $U=2$ and $U=\infty$.
\par In the Luttinger model, it is well known \cite{luttinger} that $\chi$, $u_\rho$ and $D$ 
are related through
\begin{equation}
D=2u_\rho K_\rho,
\label{ll0}
\end{equation}
with
\begin{equation}
K_\rho=\frac{\pi u_\rho}{2\chi},
\label{ll5}
\end{equation}
where $K_\rho$ is the exponent governing the decay of the correlation functions. 
In order to probe the doped region for which the lower energy spectrum of 
the $AB_2$ chain can be mapped onto the Luttinger model, we consider the ratio
\begin{equation}
\mbox{Ratio}=\frac{u_\rho}{\sqrt{D\chi/\pi}},
\label{ll1}
\end{equation}
which must be equal to one if the system is in the LL universality class 
\cite{poilblanc}. 

Since the $AB_2$ chain is not strictly 
one-dimensional, care must be taken with the length scales ($V$ and $L$) in Eqs. 
(\ref{ll2}), (\ref{ll3}) and (\ref{ll4}). For $U=0$, the orbitals at sites $A$ 
and bonding orbitals at sites $B$ are translationally equivalent and both 
build the dispersive branches shown in Fig. \ref{Disp}(c). In this case, the system can be mapped 
onto a tight-binding linear chain with $2N_c$ sites and a rescaled hopping 
parameter, $t\rightarrow t\sqrt{2}$, with $K_\rho=1$. In order that Eq. 
(\ref{ll5}) matches this result for $\epsilon_F<0$, we must choose $V\equiv L=2N_c$ with 
$\epsilon(k)=-2\sqrt{2}\cos(k)$; or, likewise, $V\equiv L=N_c$ and the dispersions 
as written in Eq. (\ref{disp}). In both cases $k_F=\frac{\pi}{2}n_0$, with 
$n_0=\frac{N_e}{L}$. Consider, for example, the former option. For $U=0$, the 
charge excitation velocity is equal to the Fermi velocity $u_F$, which can be 
easily calculated as
\begin{equation}
u_F=\left . \frac{\partial 
\epsilon(k)}{\partial 
k}\right|_{k=k_F}=2\sqrt{2}\sin(k_F).\label{ll6}
\end{equation}
On the other hand, 
substituting the GS energy,
\begin{equation}
E_{GS}(n_0)=\frac{-8\sqrt{2}}{\pi}N_c\sin(\frac{\pi}{2}n_0),
\end{equation}
into the continuous version of Eq. (\ref{ll2}), we obtain,
\begin{eqnarray}
\chi&=&\frac{1}{V}\frac{\partial^2 E_{GS}}{\partial n_0^2},\\
    &=&\pi\sqrt{2}\sin(\frac{\pi}{2}n_0) \label{ll7}.
\end{eqnarray}
Using now Eqs. (\ref{ll6}) and (\ref{ll7}) in Eq. (\ref{ll5})
we find, as expected, $K_\rho=1$. 
\begin{figure}[t]
\begin{center}
\epsfig{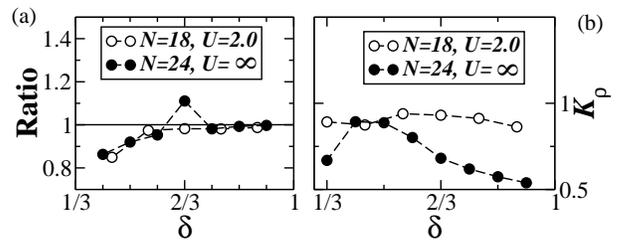}
\caption{(a) ED results for the Ratio=$u_\rho/\sqrt{D\chi/\pi}$.
(b) ED results for $K_\rho$ as function of $\delta$.}
\label{ratiokr}
\end{center}
\end{figure}
\begin{figure}[t]
\begin{center}
\epsfig{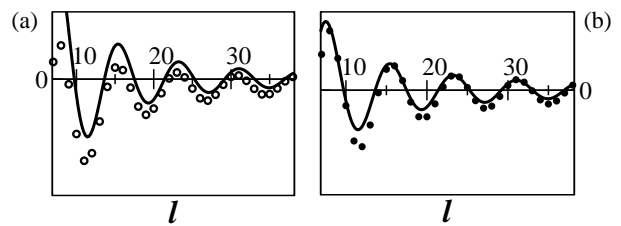}
\caption{Spin correlation functions $C(l)$ for (a) $U=2$ and (b) $U=\infty$ at 
$\delta=88/106$ for $N=106$ using DMRG: solid lines are fittings using Eq. (\ref{CLL}).}
\label{cor23dop}
\end{center}
\end{figure}

We now turn to the interacting case using ED. 
As shown in Fig. \ref{ratiokr}(a) the LL character is quite clear for 
$\delta>2/3$, while for $1/3<\delta<2/3$ we identify the crossover region. The ED 
results for $K_{\rho}$ are presented in Fig. \ref{ratiokr}(b). Notice that $K_\rho$ is close to 1
 (non-interacting fermions) for 
$U=2$; while, 
$K_\rho$ is close to $1/2$ (non-interacting {\it spinless} fermions) for $U=\infty$
\cite{luttinger,lluinf}. In order to check these results, we used DMRG to calculate the ELC spin correlation function 
\begin{equation}
C(l)\equiv\frac{\sum_{i,j}\langle\mathbf{S}_i\cdot\mathbf{S}_j\rangle\delta_{|i-
j|,l}}{\sum_{i,j}\delta_{|i-j|,l}},
\label{dmrgcorr}
\end{equation}
whose asymptotic behavior should match that for the Luttinger model \cite{lluinf}:
\begin{equation}
C_{LL}(l)\sim \frac{\cos(2k_F l)[\ln(l)]^{1/2}}{l^{1+K_\rho}}.
\label{CLL}
\end{equation}
In Eq. (\ref{dmrgcorr}) we have considered an average over all possible pairs of sites
separated by the same distance $l$, a procedure that reduces open boundary effects. 
In Figs. \ref{cor23dop}(a) and \ref{cor23dop}(b)
we show $C(l)$ calculated at $\delta=88/106$ for $U=2$ and $U=\infty$,
respectively. Also shown are the fittings to $C(l)$ using $C_{LL}(l)$ with 
$k_F=\frac{\pi}{2}n_0$ and $K_\rho$ taken from the results shown in Fig. 
\ref{ratiokr}(b) after linear interpolation: $K_\rho=0.89$ ($U=2$) and 
$K_\rho=0.57$ ($U=\infty$). Motivated by a compromise between large values of $l$ and minimum boundary effects, 
we have considered intermediate values of $l$ in the fitting,
which is quite good for both values of $U$. We thus conclude that the Luttinger model correctly describes 
the low energy physics of the $AB_2$ chain for $\delta>2/3$.
\section{Summary and Conclusions}

In summary, the numerical results presented 
here have clearly evidenced the rich phase diagram
exhibited by the Hubbard model on the doped AB$_2$ chain both for $U=2$ and in 
the infinite-U limit. We have shown that at the commensurate dopings $\delta=1/3$ and $2/3$ the system display 
insulating phases, although for $U=2$ the charge gap $\Delta_c$ is very small at $\delta=2/3$, with indications
that $\Delta_c$ present an essential singularity as $U\rightarrow 0$. For $U=2$ and $\delta \lesssim 0.02$ the
GS exhibit a ferrimagnetic phase reminiscent of the undoped regime, while for $0.02 \lesssim \delta \lesssim 0.07$
incommensurate magnetic correlations are observed. For $U=\infty$ and $\delta=0$ the GS total spin is degenerate,
 whereas for $0<\delta \lesssim 0.225$ hole itinerancy (Nagaoka mechanism) sets a fully polarized GS.
In this case, we have also observed the presence of an extensive number of low-lying levels with total
 spin ranging from the minimum value to $S_{max}-1$ and level spacing decaying with system size as $1/N_c$.
  For higher doping, the system phase separates into coexisting metallic and insulating
 phases for $\delta_{PS}(U)\lesssim \delta <1/3$ (with $\delta_{PS}(\infty)\approx 0.225$ and 
 $\delta_{PS}(2)\approx 0.07$). The insulating state presents a finite spin gap and fully fills the
  system at $\delta=1/3$, which is well described by a short-ranged-RVB state. Finally, a crossover region
  is observed for $1/3<\delta <2/3$, while a Luttinger liquid behavior is explicitly characterized
  for $\delta>2/3$. 

In closing, we would like to stress that the above-reported results might also stimulate further experimental and theoretical 
investigations on quasi-one-dimensional compounds displaying
complex unit cell structures \cite{batista}.

We acknowledge useful discussions with A. L. Malvezzi and M. H. Oliveira.
This work was supported by CNPq, Finep, FACEPE and CAPES (Brazilian agencies).

\appendix
\begin{figure}[b]
\begin{center}
\epsfig{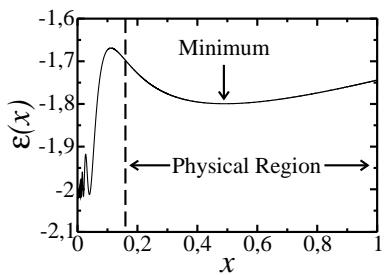}
\caption{Energy per unit cell $\epsilon$ as function of 
the fraction $x=L_-/N_c$ for $\delta=0.28$. The physical
region is also shown.}
\label{Var}
\end{center}
\end{figure}
\section{Variational Approach for $U=\infty$ and $\delta\leq 1/3$}
In the metallic saturated ferromagnetic region (parity symmetry -1) the energy as function of
doping is known to have a non-interacting spinless fermion behavior:
\begin{equation}
E(k_{F-})=-\frac{4\sqrt{2}}{\pi}L_-\sin(k_{F-}/2),
\end{equation}
where $k_{F-}=\pi \nu_{h-}$, $\nu_{h-}=N_{h-}/L_{-}$, and
$L_-$ is the linear size of the system. 
On the other hand, in the insulating paramagnetic phase 
(SR-RVB states with even parity symmetry) at $\delta=1/3$ (one hole per cell)
\begin{equation}
N_{h+}=L_+,
\label{var_1}
\end{equation}
and the energy per cell $\epsilon_+$ is almost independent of the system linear 
size and can be estimated either by using ED or DMRG:  
\begin{equation}
\epsilon_{+}\approx -2.021.
\end{equation}

Let us now consider a phase-separated regime in which a paramagnetic phase
with size $L_+$ coexists with a ferromagnetic one with size $L_-$, so the
energy per cell reads 
\begin{equation}
\epsilon=\epsilon_+ 
\frac{L_+}{N_c}-\frac{4\sqrt{2}}{\pi}\frac{L_-}{N_c}\sin(\frac{\pi}{2}\nu_{h-}).
\label{var1}
\end{equation}
It is convenient to write $\nu_{h-}$ as
\begin{equation}
\nu_{h-}=\frac{N_h-(N_c-L_-)}{L_-}=\frac{3\delta-(1-x)}{x}.
\label{var0}
\end{equation}
where $N_h=N_{h+}+N_{h-}$, $N_c=L_++L_-$, $x=L_-/N_c$ and $N=3N_c$.
Using the above notation we rewrite Eq. (\ref{var1}) in the form below 
\begin{equation}
\epsilon(x)=(1-x)\epsilon_+-\frac{4\sqrt{2}}{\pi}x\sin\left[\frac{\pi}{2}\left(
\frac{3\delta-1}{x}+1\right)\right]\label{var4}
\end{equation}
Here we should notice the presence of a singularity at $x=0$ for any finite 
value of 
$\delta\neq 1/3$ (see Fig. \ref{Var}). However, the region of physical 
values of $x$ is defined by
\begin{equation}
0\leq N_{h+}\leq N_h, \mbox{ i. e.,}
\end{equation}
\begin{equation}
1-3\delta \leq x \leq 1.
\label{var_2}
\end{equation}
In Fig. \ref{Var} we present $\epsilon(x)$ for $\delta=0.28$, in which 
the physical region is $0.16\leq x \leq 1$ and can be found by Eq. 
(\ref{var_2}), with a minimum in $\epsilon(x)$ for $x\approx 
0.49$. 

The value of $x$ which minimizes the energy for a given $\delta$, 
$\overline{x}=\overline{x}(\delta)$, satisfies the equation 
$\left[\frac{\partial \epsilon(x)}{\partial x}\right]_\delta=0$, which can be 
written as
\begin{equation}
\frac{\pi\epsilon_+}{4\sqrt{2}}=\cos(y)+y\sin(y),
\label{var5}
\end{equation}
where
\begin{equation}
y\equiv\frac{\pi}{2}\frac{3\delta-1}{\overline{x}}.
\end{equation}
The roots of Eq. (\ref{var5}) are numerically calculated and conduct to 
\begin{equation}
\left \{
\begin{array}{lll}
\overline{x}=1,&\mbox{for }&\delta\lesssim 0.225;\\
\overline{x}\approx 3.071-9.213\delta,&\mbox{for }& 0.225\lesssim \delta\leq \frac{1}{3}.
\end{array}
\right.
\label{var11}
\end{equation}
We thus conclude that $\delta_{PS}(\infty)\approx 0.225$, which is in very good
agreement with ED and DMRG calculations.
\par The magnetization is null at the even phase and maximum at the odd one.
We can thus derive the following expression for the GS total spin per unit cell:
\begin{eqnarray}
\frac{S_g}{N_c}&=&\frac{1}{2N_c}(N_e-2L_+)\\
               &=&\frac{1}{2}[3(1-\delta)-2(1-\overline{x})]
\end{eqnarray}
The dependence of the average parity $p$ on $\delta$ can also be easily 
written as
\begin{equation}
p=1-2\overline{x}
\end{equation}
Finally, using Eq. (\ref{var11}) for $\overline{x}$,
the above results for $p$ and $S_g$ are plotted in Figs. \ref{pdEz}(d) and 
\ref{sqsefl}(c), respectively, and shown to be in excellent agreement with the ED and DMRG
calculations.

\end{document}